\documentclass{vgtc}                          

\ifpdf
  \pdfoutput=1\relax                   
  \usepackage{graphicx}                
  \DeclareGraphicsExtensions{.pdf,.png,.jpg,.jpeg} 
\else
  \usepackage{graphicx}                
  \DeclareGraphicsExtensions{.eps}     
\fi%

\graphicspath{{figures/}{pictures/}{images/}{./}} 

\usepackage{microtype}                 
\PassOptionsToPackage{warn}{textcomp}  
\usepackage{textcomp}                  
\usepackage{mathptmx}                  
\usepackage{times}                     
\usepackage{cite}                      
\usepackage{tabu}                      
\usepackage{booktabs}                  

\onlineid{0}

\vgtccategory{Research}

\vgtcinsertpkg

\title{Visualizing Gender Gap in Film Industry over the Past 100 Years}

\author{Junkai Man %
\and Ruitian Wu%
\and Chenglin Zhang 
\and Xin Tong\thanks{Corresponding Author: Xin Tong -- Division of Natural and Applied Sciences, Duke Kunshan University, Kunshan, Jiangsu 215316, China; Email: xt43@duke.edu. Authors: Junkai Man, Ruitian Wu, Chenglin Zhang -- Division of Natural and Applied Sciences, Duke Kunshan University, Kunshan, Jiangsu 215316. This report is co-authored by the three authors.}}
\affiliation{\scriptsize Duke Kunshan University}

\abstract{Visualizing big data can provide valuable insights into social science research. In this project, we focused on visualizing the potential gender gap in the global film industry over the past 100 years. We profiled the differences both for the actors/actresses and male/female movie audiences and analyzed the IMDb data of the most popular 10,000 movies (the composition and importance of casts of different genders, the cooperation network of the actors/actresses, the movie genres, the movie descriptions, etc.) and audience ratings (the differences between male’s and female’s ratings). Findings suggest that the gender gap has been distinct in many aspects, but a recent trend is that this gap narrows down and women are gaining discursive power in the film industry. Our study presented rich data, vivid illustrations, and novel perspectives that can serve as the foundation for further studies on related topics and their social implications.%
} 

\CCScatlist{
  \CCScatTwelve{Data}{Visualization}{Movie Industry}{Genders};
}

\begin{document}


\firstsection{Introduction}

\maketitle

\begin{figure}[htbp]
 \centering 
 \includegraphics[width=0.8\columnwidth]{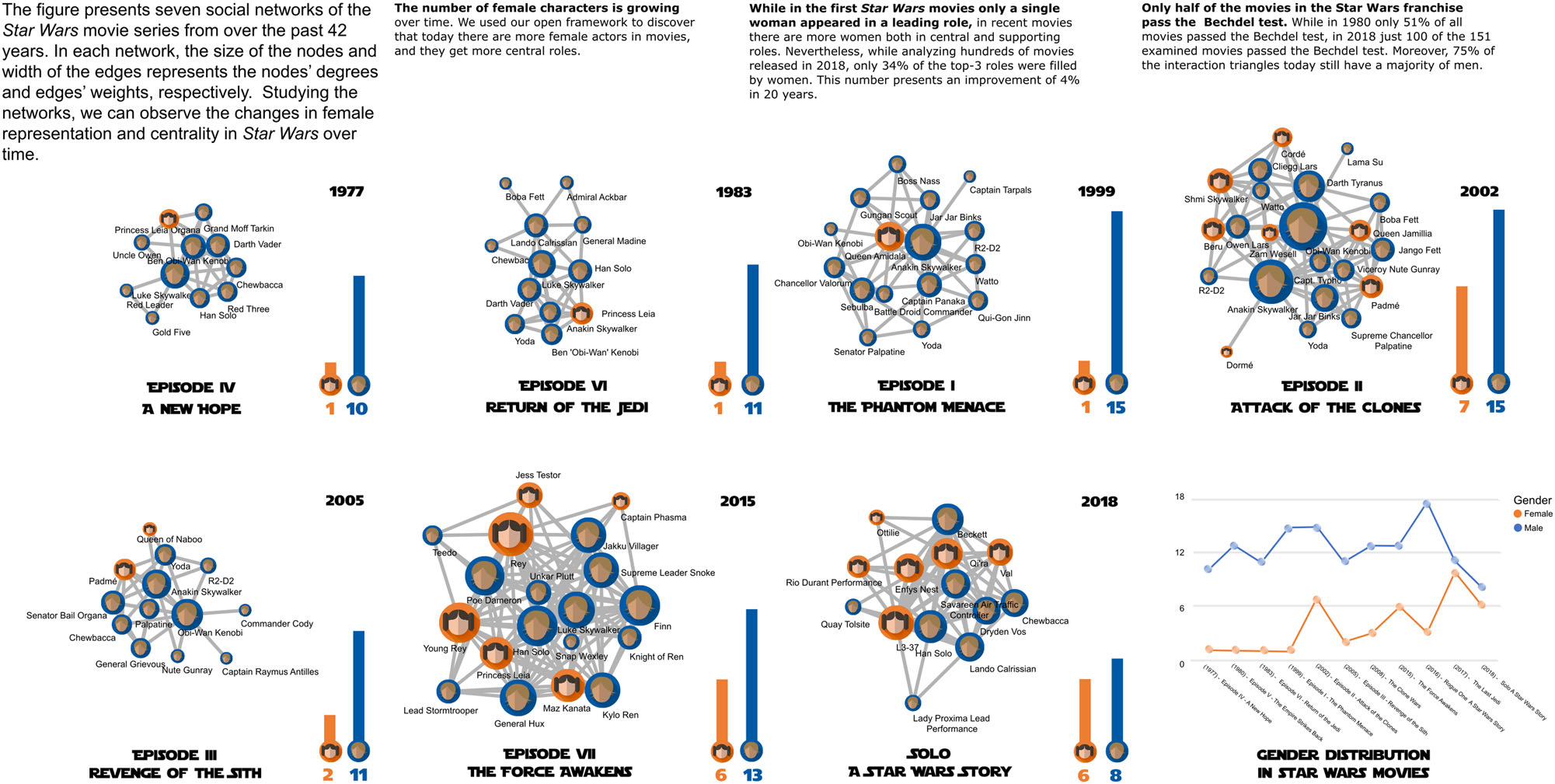}
 \caption{The evolution of female representation in the Star Wars movies series, retrieved from \cite{DBLP:journals/corr/abs-1903-06469}.}
 \label{Figure 1}
\end{figure}

\indent The colorful and epic motion picture industry has enjoyed more than 100 years of prosperity; at the same time, it reflects social trends, public opinions, and even stereotypes. In turn, great movies and producers also profoundly impact on the development of the movie industry and society itself. We are inspired by the famous movie series Star Wars, which is mentioned by Kagan et al. \cite{DBLP:journals/corr/abs-1903-06469} in Figure \ref {Figure 1}. The gender ratio in the first few Star Wars episodes is quite skewed, but it's more balanced than before for most recent episodes. In our work, we want to investigate this phenomenon of the merging gender gap in a broader context. In addition, we also want to study the changes in the preferences of male and female audiences since the audience ratings of the recent Star Wars episodes are lower than the early ones.

Males and females play different roles either as actors/actresses in the movie production process or as audiences. In the movie production process, actors' and actresses' engagement differs in movie ratings (e.g., PG-13, R), genres, movie content, and cooperative networks. As audiences, males' and females' movie preferences vary over time, e.g., their ratings for each movie. Therefore, we define the male and female's differences in these two aspects as gender gaps. In this project, we aim to visualize this gender gap using time-series data from 1920 to 2021 to (1) prove the existence of this gender gap and (2) further understand its evolution process over the past 100 years. By looking into these visualizations, we can gain valuable insights into the general trend of the gender gap over the 100 years, and thus will have a starting point for further investigation and statistical examination.

More precisely, our visualizations aim to: 1) Compare actor- and actress-dominant movie genres for the past 100 years. 2) Identify collaborations among actors and actresses. 3) Categorize the yearly number of films by age ratings and protagonist genders. 4) Compare the descriptions for the actor- and actress-dominant movies. 5) Identify male and female audiences' movie preferences.

\section{Visualization Method and Results}
Here, we introduce the data collection, processing, and visualization methods to achieve our research goals. In total, we crawled 10,200 movie items from 1920 to 2021, with 100 most voted movies each year. Each data item includes the movie title, year, genre, runtime, certificate, IMDb rating (total and for male and female raters), movie description, director name, star name, their corresponding IMDb ID, number of votes, and box office takings (if available). Moreover, we calculated an index to indicate whether the movie is actor- or actress-dominant by checking the cast list and the sequence of the stars listed in it. Then, we split the dataset into two groups (actor and actress-dominant) by the index, which is about 3:2 in size (actor: actress).

\subsection{Co-stardom Network}
First, we compared how actors and actresses cooperate differently in different periods through a network diagram, showing times of co-occurrences between actors/actresses. In Figure 2, nodes represent actors/actresses while edges represent co-occurrences between different actors/actresses. Moreover, colors represent genders (yellow as male and blue as female), and the size refers to the times of occurrences for each actor/actress. Although we can observe that actors make up the main skeleton of the whole network and occupy the most significant nodes, actresses are playing more and more important roles in the network as time passes (Figure 2). The network is becoming more balanced in terms of gender, which indicates that actresses are securing a relatively equal place in the film industry compared to actors. Such balance can be reflected by movie ratings as well. For example, in recent years, the rating distribution is becoming more homogeneous, indicating that the gap between the two types of movies is decreasing (they are shifting downward). Nowadays, restrictions on ratings of movies that actresses could star in are being gradually removed. 

\begin{figure}[htbp]
 \centering 
 \includegraphics[width=\columnwidth]{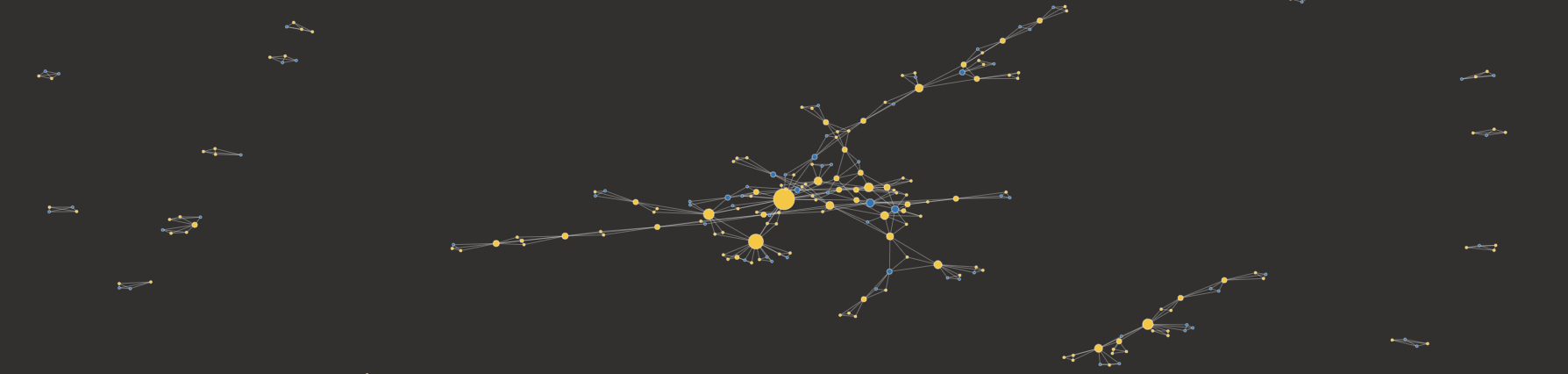}
 \caption{Co-stardom Network}
 \label{fig:1}
\end{figure} 

\subsection{Movie Rating Theme River}
Rating of movies (e.g., PG-13, R, etc.) is an important characteristic featuring the target audience of the movies. We visualized a Theme River diagram to show the distribution of rating differences between actor- and actress-dominant movies. Figure 3 places actor- and actress-dominant movies above/below the x-axis separately. Colors were used to represent ratings, and their widths represent times of occurrences. In this visualization, types of ratings could be selected in a customized way. 

\begin{figure}[htbp]
 \centering 
 \includegraphics[width=\columnwidth]{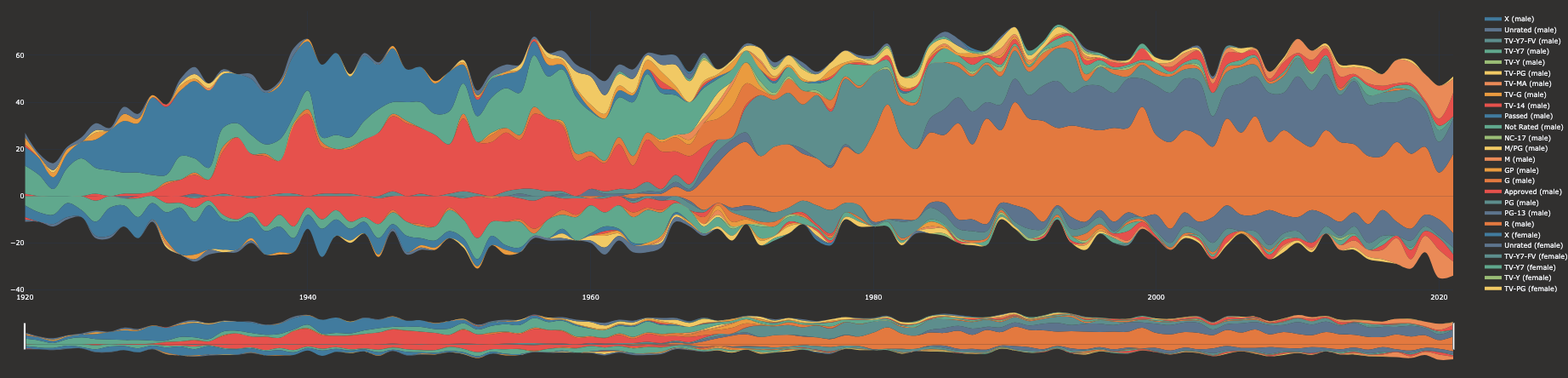}
 \caption{Movie Rating Theme River}
 \label{fig:1}
\end{figure} 

\subsection{Movie Theme Word Cloud}
Another important indicator for movies is its keyword, so we extracted keywords from the movies' descriptions. We visualized two word clouds in Figure 4 to illustrate how actor- and actress-dominant movies' keywords differ. Clear distinctions could be observed between two genders, even though some similar general topics were shared by the two groups. For actor-dominant ones, there are also topics related to "war", "world" and "murder”. For actress-dominant ones, some words like "girl", "wife" and "mother" are observed, indicating more explorations of self-identity in these movies. The main story behind actor- and actress-dominant movies still varies, reflecting of diversity instead of the gender gap.


\begin{figure}[htbp]
 \centering 
 \includegraphics[width=\columnwidth]{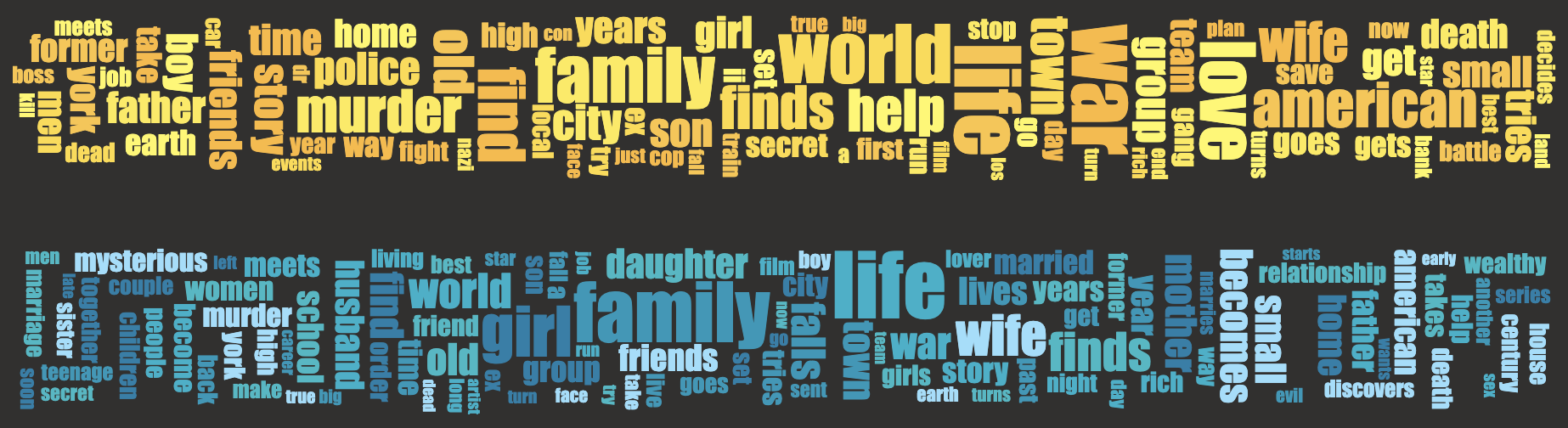}
 \caption{Movie Theme Word Cloud}
 \label{fig:1}
\end{figure} 

\subsection{Genre Chord Diagram}
 We developed a Chord Diagram (Figure 5) to compare the genre differences between actor- and actress-dominant movies. In Figure 5, the left Chord represents the distribution of film genres for actor-dominant movies. In contrast, the right Chord represents the distribution of film genres for actress-dominant movies. In the visualization, a time period could be selected so that users can observe the changes over time. We found that before the 1970th, most actresses starred in drama and romance films while actors dominated the movies of action, crime, and adventures. However, the gap has been narrowing in recent years as the composition of two chord diagrams are approaching the same. When choosing actors/actresses to star in a particular film, gender has become a less critical indicator to rely on.

\begin{figure}[htbp]
 \centering 
 \includegraphics[width=0.7\columnwidth]{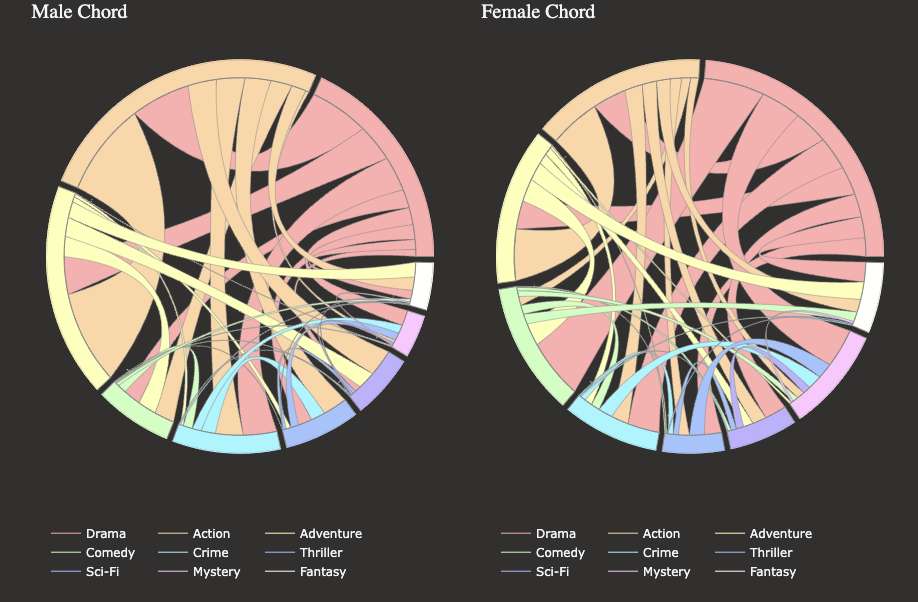}
 \caption{Genre Chord Diagram}
 \label{fig:1}
\end{figure} 

\subsection{Bubble Chart of Audience Preferences}
Besides actors/actresses' gender gap, the audience is also a crucial part of considering when analyzing the gender gap. Here, we presented two bubble charts showing similarities and differences in audience preference for the two genders. In Figure 6, the top figure shows aggregated ratings for both male and female audiences, while the one on the bottom shows their preference gap. A higher value would indicate a higher preference for male audiences, and vice versa. While the gender gap between actors/actresses is being narrowed down, the audience gap is evolving in the opposite direction. We noticed that compared to movies made before 2000, female audiences have an equal or even higher weight in determining the popularity of movies. The gap also indicates that the gap in the taste between male and female audiences is becoming larger than before. Producers nowadays should spend more efforts to satisfy audiences of different genders.

\begin{figure}[htbp]
 \centering 
 \includegraphics[width=\columnwidth]{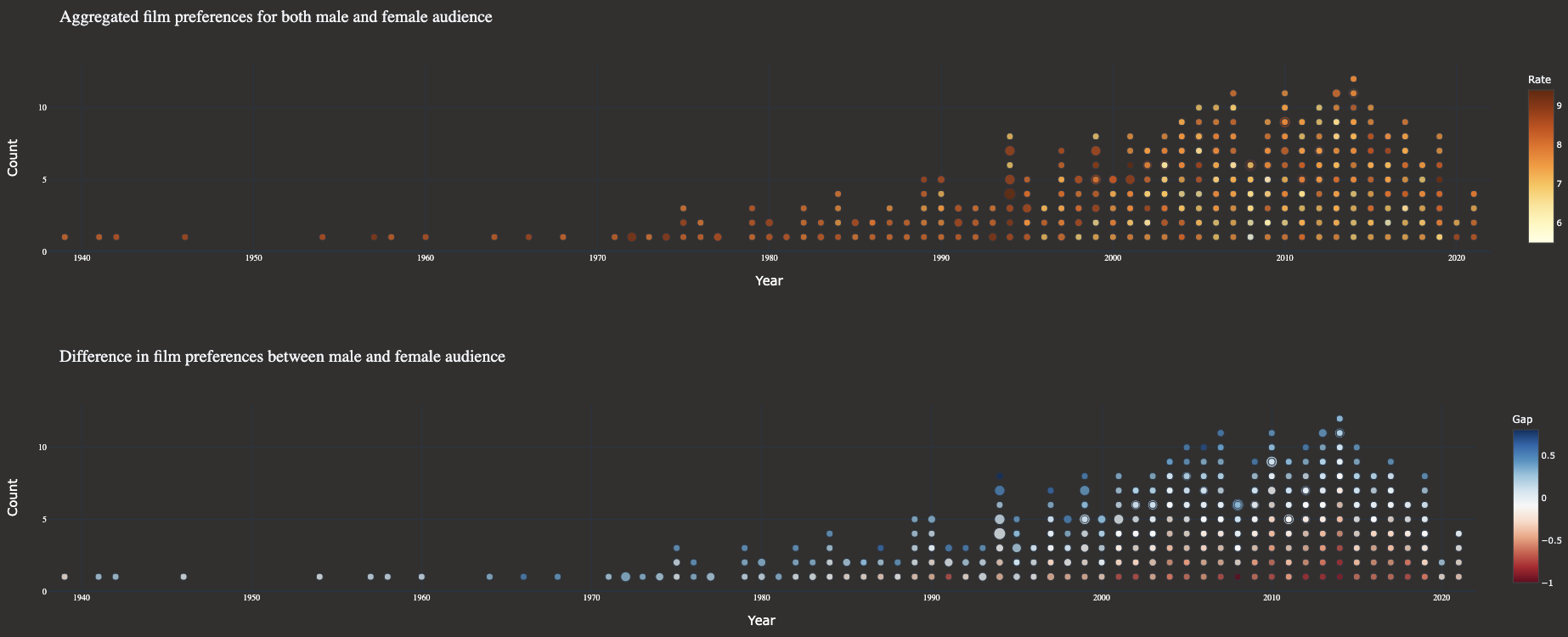}
 \caption{Audience Preference Bubble Chart}
 \label{fig:1}
\end{figure} 

\section{Conclusion}
In conclusion, from our visualizations, findings suggest that the gender gap in the movie industry is diminishing, especially in movie casts. However, the preferences of male and female audiences have become more diverse in recent years. The five visualizations give us a glimpse into movie industry's development  and potential trends in the future, which is essential when more people are advocating for Feminism. Overall, this visualization project provides researchers with foundations to dig into this gender gap topic in movies.\footnote{The demo page of this project can be found at\\ https://junkaiman.com/projects/gender-gap-in-film}
\acknowledgments{
The authors thank Duke Kunshan University for the academic support throughout this project.}
\bibliographystyle{abbrv-doi}

\bibliography{template}
\end{document}